\documentclass[
    sigconf
]{acmart}

\copyrightyear{2024}
\acmYear{2024}

\setcopyright{rightsretained}
\acmConference[PASC '24]{Platform for Advanced Scientific Computing Conference}{June 3--5, 2024}{Zurich, Switzerland}
\acmBooktitle{Platform for Advanced Scientific Computing Conference (PASC '24), June 3--5, 2024, Zurich, Switzerland}%

\acmDOI{XXXXXXX.XXXXXXX}

\usepackage[utf8x]{inputenc}

\begin{document}

\title{Synthesizing Particle-in-Cell Simulations Through Learning and GPU Computing for Hybrid Particle Accelerator Beamlines}

\author{Ryan T. Sandberg}
\email{{rsandberg,axelhuebl}@lbl.gov}
\orcid{0000-0001-7680-8733}
\author{Remi Lehe}
\orcid{0000-0002-3656-9659}
\author{Chad E. Mitchell}
\orcid{0000-0002-1986-9852}
\author{Marco Garten}
\orcid{0000-0001-6994-2475}
\author{Andrew Myers}
\orcid{0000-0001-8427-8330}
\author{Ji Qiang}
\orcid{0000-0002-2537-483X}
\author{Jean-Luc Vay}
\orcid{0000-0002-0040-799X}
\author{Axel Huebl}
\orcid{0000-0003-1943-7141}
\affiliation{%
  \institution{Lawrence Berkeley National Laboratory}
  \streetaddress{1 Cyclotron Rd}
  \city{Berkeley}
  \state{California}
  \country{USA}
  \postcode{94720}
}


\begin{abstract}
Particle accelerator modeling is an important field of research and development, essential to investigating, designing and operating some of the most complex scientific devices ever built.
Kinetic simulations of relativistic, charged particle beams and advanced plasma accelerator elements are often performed with high-fidelity particle-in-cell simulations, some of which fill the largest GPU supercomputers.
Start-to-end modeling of a particle accelerator includes many elements and it is desirable to integrate and model advanced accelerator elements fast, in effective models.
Traditionally, analytical and reduced-physics models fill this role.
The vast data from high-fidelity simulations and power of GPU-accelerated computation open a new opportunity to complement traditional modeling without approximations: surrogate modeling through machine learning.
In this paper, we implement, present and benchmark such a data-driven workflow, synthesising a fully GPU-accelerated, conventional-surrogate simulation for hybrid particle accelerator beamlines.
\end{abstract}




\received{1 December 2023}
\received[revised]{23 February 2024}
\received[accepted]{25 April 2024}

\settopmatter{printfolios=true}

\maketitle

\section{Introduction\label{sec:intro}}
Particle accelerators are among the most complex scientific devices ever built, used in medical and industrial applications or the frontier of high-energy physics.
Particle accelerator facilities are modeled extensively with various scientific codes, many of which use specialized algorithms based on the particle-in-cell method~\cite{BiedronSM2022,VayJoI2021}.
Thus, high-performance simulations are a pillar of modern particle accelerator design.

Kinetic simulations of particle accelerators in varying levels of detail and approximations are often necessary to capture the full dynamics at play.
The authors of this paper, together with their collaborators, develop the open source Beam pLasma \& Accelerator Simulation Toolkit (BLAST).\footnote{\url{https://blast.lbl.gov}}
Two central codes in BLAST for this paper are \verb+WarpX+~\cite{FedeliHuebl2022}, a time-evolving particle-in-cell code that was awarded the 2022 Gordon Bell Prize in high-performance computing, and the recently published code \verb+ImpactX+, a beam dynamics code evolving a particle bunch relative to a reference trajectory~\cite{HueblNAPAC22}.\footnote{\url{https://github.com/ECP-WarpX}}
Complemented by \verb+HiPACE+++~\cite{Diederichs2022}, a specialized code for plasma-wakefield modeling, these three codes started a new ``Exascale-era'' set of multi-node, fully GPU-accelerated and mesh-refinement capable codes in BLAST~\cite{HueblAAC22}.
These BLAST codes share common HPC dependencies for data distribution and performance portability through the \verb+AMReX+ library~\cite{AMReX_JOSS}.
More generally, BLAST codes reuse further shared middleware libraries and API conventions, and I/O data compatibility through the open particle-mesh metadata standard (openPMD)~\cite{openPMDstandard}.

The value of the full BLAST suite is that it covers modeling needs on different time and length scales, as well as different levels of fidelity and time-to-solution, in a single framework of interoperable codes~\cite{HueblAAC22}.
Since modeling the detailed physics can be expensive, there is an interest to capture specialized scenarios with reduced-order and surrogate models.
Typical examples of reduced-order models include reduction of fidelity, neglecting electromagnetic effects, assuming low energy spreads or limiting the geometry to (quasi-)axisymmetric scenarios.

For start-to-end (whole-device) modeling in particle accelerator design and research, reference trajectory-based modeling is very efficient in describing bunch transport and dynamics evolution over long distances, often over meters to kilometers.
In combination with using the accelerator lattice position $s$ as the independent dynamical variable, particle phase space coordinates are specified relative to a nominal reference trajectory and particles may only deviate slightly from it~\cite{LeeBook,Qiang2000}.
More general, time-based modeling uses the time $t$ as the independent dynamical variable without further assumptions on the particle distribution.
This can be more costly, but enables modeling of scenarios with complex particle trajectories, as are present in advanced accelerator elements that are based on plasma elements for acceleration~\cite{Esarey2009}.




In this work, we explore a data-driven approach to fast modeling of specialized beamline elements.
When collective effects are negligible, particle accelerator elements can be described as complex transfer maps for each particle, $f:\mathbb{R}^6\to\mathbb{R}^6$, mapping initial 6D particle phase space coordinates to final phase space coordinates.
One can deploy machine-learning surrogates such as neural networks $F_{nn}$ to approximate these transfer maps to predict particle phase space coordinates coming out of the accelerator elements as functions of phase space coordinates at the entrance to the accelerator element.
This formulation is based on the assumption that the beam particles do not interact and is only valid when there are no collective effects (i.e., no beam space-charge, wakefield, synchrotron radiation, etc.), as is the case for traditional transfer maps~\cite{LeeBook}.

\section{Synthesis of Data and Compute: Bridging Scales Through Machine-Learning}
\label{sec:impactx_and_surrogates}

In this paper, we employ data and compute models in a combined simulation for fast accelerator modeling.
We refer to this as a synthesis, combining traditional physics-based simulations in high-performance computing in situ with data-driven models that can approximate computationally costly sections of an accelerator beamline.
In particular, we combine:
(i) a fast modeling approach to study beamline hybrids that includes conventional and advanced, non-linear accelerator elements and
(ii) a machine-learning augmented high-performance simulation.

For the first part, we combine advanced accelerator elements (e.g., plasma-based) with conventional beamline elements.
For instance, plasma elements such as laser-wakefields and plasma lenses are often short (tens of centimeters) and can require high-fidelity modeling of often many 1,000s of GPU hours for predictive quality.
Conventional beamline elements on the other hand are fast in modeling, but span large scales ($\gg$ meters).
Typically, a modeling workflow for advanced accelerator elements would involve a $t$-based code such as \verb+WarpX+ for the detailed, plasma-based elements and an $s$-based code such as \verb+ImpactX+ for long-scale transport sections, with communication between the two codes done via file I/O of the particle bunch data.

For the second part, we aim to create a fast simulation with \verb+ImpactX+ that integrates the usually \verb+WarpX+-modeled, costly, advanced accelerator elements at a computational cost comparable to conventional tracking costs in $s$-based codes.
For this, the strategy in this work incorporates machine-learning surrogates into GPU-accelerated simulations using node-local inference.
Machine-learning models themselves are created from high-fidelity data from BLAST HPC PIC simulations - training a transfer-map like model reduces these simulation results to a specialized machine-learning data model.

Aiming for a seamless integration of HPC modeling codes and machine-learning ecosystems, we extended both \verb+AMReX+-based, GPU-accelerated codes \verb+WarpX+ and \verb+ImpactX+ with Python bindings using the new \verb+pyAMReX+ library~\cite{pyAMReX,AMReXIJHPCA2024}.
\verb+pyAMReX+ implements emerging community standards for zero-copy, in-memory data exchange such as the (CUDA) Array Interface~\cite{arrayinterface} or alternatively the open in-memory tensor structure (DLPack) for sharing arbitrary data among frameworks.
Through this adherence to community standards, one can create non-owning, read-write enabled \textit{views} of node-local, HPC simulation data to Python frameworks such as \verb+numpy+ on CPUs, \verb+cupy+ on GPUs, permitting to add at runtime custom, just-in-time compiled GPU kernels, or even GPU-accelerated ML frameworks such as \verb+PyTorch+~\cite{pytorch}.

Python scripting layer and data interfaces enable \verb+WarpX+ and \verb+ImpactX+ users to design and augment their fully (GPU) accelerated \verb+AMReX+ C++-based simulations productively.
\verb+ImpactX+ implements callback functions from C++ into Python code that can be adjusted at runtime to program beam optics elements from Python.
The callbacks can advance the beam particles through any of three different approaches: 
\begin{enumerate}
    \item use the Python zero-copy data views to implement additional transfer maps and integrators as in traditional accelerator modeling;
    \item call back into accelerated C++ code;
    \item advance their particle tensors through inference with a pre-trained machine-learning surrogate model, thus completing the high-performance compute model model synthesis with data modeling.
\end{enumerate}
The presented approach works on CPU-only hardware too, but since it is desirable to accelerate machine-learning surrogate inference \textit{and} traditional modeling components, this work goes one step further.
Both the traditional HPC simulation compute model and the machine-learning data model are fully GPU-accelerated, without involving a host-device particle copy, enabling significant performance benefits over existing approaches~\cite{Badiali2022}.

For particle beamline element tracking, the in situ inference of ML models is very fast compared to the traditional approach of switching codes through I/O and can be done node-local on each subset of distributed particles in \verb+ImpactX+, which simplifies model deployment in multi-node parallel execution, as is typical for HPC simulation runs.

\section{Related Work: Neural Networks as fast Surrogates in Particle-in-Cell Modeling\label{sec:neural_network_surrogates}}

Neural networks are a class of machine learning algorithms that have been explored as versatile and robust in learning complex functions, and so serve as surrogate models~\cite{deeplearning,deeplearning_overview}.
In particle-in-cell modeling for accelerator and laser-plasma science, examples include:

\begin{itemize}
    \item Koser et al.~\cite{Koser2022} used neural networks as surrogate models for optimizing RF quadrupole accelerator design.
    \item Djordjevic et al.~\cite{Djordjevic2021} trained a feed-forward neural network surrogate from low-dimensional PIC simulations for predicting beam properties in laser-ion acceleration.
    \item Badiali et al.~\cite{Badiali2022}, augmented CPU-only Fortran PIC code for collisional physics with limited speedup potential and surrogate precision.
    \item Edelen et al.~\cite{EdelenPRAB2020} used a neural network trained with PIC simulations in order to accelerate the design optimization of a particle accelerator.
    \item In a recent conference proceedings~\cite{SandbergIPAC23}, we reported our progress modeling accelerator elements with neural networks.  In this work we have improved on the neural network structure and training, trained on 15 stages, and incorporated the neural networks as surrogates into a larger accelerator modeling workflow.
\end{itemize}

A neural network is a parametric nonlinear function $F_{NN}:\mathbb{R}^n\to\mathbb{R}^m$ whose parameters can be adjusted to approximate another function $f:\mathbb{R}^n\to\mathbb{R}^m$. 
The structure of the neural network and the approximation process are described by more parameters, commonly called hyperparameters.
A fully connected feed-forward neural network has the structure of many interleaved compositions of simple nonlinear functions with linear connections. 
The parameters of the neural network are called its weights, denoted collectively here as the matrices $\{\mathbf{A_i}\}_{i=0,\ldots,\ell}$ and biases, denoted collectively here as $\{\vec b_i\}_{i=0,\ldots,\ell}$. 
Here $\ell\in\mathbb{N}$ is the number of hidden layers of the neural network.
Let $\mathbf{A_0}\in \mathbb{R}^{H_1\times m}$, $\mathbf{A_i}\in \mathbb{R}^{H_{i+1}\times H_{i}}$ for $i=1,\ldots,\ell-1$, and $\mathbf{A_\ell}\in\mathbb{R}^{m\times H_{\ell}}$, $b_i\in \mathbb{R}^{H_{i+1}}$ for $i=0,1,\ldots,\ell-1$, and $\vec b_\ell\in\mathbb{R}^{m}.$ 
Here $H_i\in\mathbb{N}$ is the number of nodes per hidden layer of the neural network. 

The ability of the neural network to approximate nonlinear functions comes from its activation functions $\sigma_i:\mathbb{R}^{H_i}\to\mathbb{R}^{H_i}$ for $i=1,\ldots,\ell$, typically the component-wise application of a simple nonlinear function such as a sigmoidal function or the Rectified Linear Unit (ReLU) piecewise function
\begin{equation}
\mathrm{ReLU}(x)=\begin{Bmatrix} 0 & x \le 0\\ x & x\ge 0\end{Bmatrix}.
\end{equation}
In this work, we use a generalization of the ReLU activation function, the Parametric Rectified Linear Unit (PReLU)~\cite{He2015}.
The PReLU activation function has a slight nonzero slope $a$ for negative values:
\begin{equation}
\mathrm{PReLU}(x)=\begin{Bmatrix} ax & x \le 0\\ x & x\ge 0\end{Bmatrix}.
\end{equation}
This slope $a$ is a model parameter to be learned along with the model weights and biases.
The PReLU activation function helps avoid dead neurons -- neurons that always output zero due to very nonnegative weights or biases preceding the neuron.
Define $L_i:\mathbb{R}^{H_{i}}\to\mathbb{R}^{H_{i+1}}$ by $L_i(\vec x)=\mathbf A_i\vec x+\vec b_i$. 
The action of the neural network can be written as 
\begin{equation}
    F(\vec x)=L_\ell\circ \sigma_\ell\circ\ldots\circ L_1\circ \sigma_1\circ L_0.
\end{equation}
In this work, we use the same activation function $\sigma$ at each layer, so $\sigma_i=\sigma$ for all $i=1,\ldots,\ell-1$.  
The hidden layers all have the same size $H$, so $H_i=H$ for all $i=1,\ldots,\ell$. 
We also look for maps where $n=m$, that is we are approximating functions $f:\mathbb{R}^n\to\mathbb{R}^n$.
In this work we did not investigate unevenly sized hidden layers.
We choose $n=6$ for the six dimensional phase space coordinates $(\vec r, \vec p)^\top$ of each particle in a bunch.

There are several results proving the ability of neural networks to approximate continuous functions, differing in the assumed structure of the network, e.g.~\cite{Hornik1989,Park2021,Shen2022}.  
While these theorems provide some assurance that in various limits of network size, and even for finite sized neural networks, parameters can be found to approximate a function arbitrarily well, they do not indicate how to obtain the parameters. 
Commonly, the neural network parameters are obtained by phrasing the approximation problem as an optimization problem -- minimizing the value of some loss function calculating the difference between the neural network prediction $\hat y_j=F(\vec x_j)$ and the ground truth $y_j=f(\vec x_j)$ for a dataset of $N$ known pairs $\{x_j,y_j=f(x_j))\}_{j=1,\ldots,N}.$  
The neural network ``learns'' or is ``trained'' as the optimization problem is solved and the optimal weights and biases are found that minimize the error of the neural network.

In this work, we use the mean-squared-error (MSE) loss function,
\begin{equation}
{\rm MSE}(\vec x, \vec y) = \frac{1}{Nn} \sum_{i=1}^N\sum_{j=1}^n (x_{ij}-y_{ij})^2,    
\end{equation}
where $\vec x, \vec y\in\mathbb{R}^{N\times n}$. 
The loss function is evaluated on the entire training set.

Common practice is to solve the optimization problem via some flavor of stochastic gradient descent.  
We use the ADAM algorithm~\cite{adam}, an adaptive stochastic gradient descent method. 
This allows for performing gradient descent over subsets of the available data, called batches, allowing for faster optimization. 
There are several possible hyperparameters associated with the optimization.  
These include the step size or learning rate of each gradient descent step $\alpha$, the number of optimization steps, often referred to as epochs, and the size of the batches.

Determining the optimal hyperparameters, number of hidden layers $\ell$, number of nodes per hidden layer $H$, activation function $\sigma$, batch size, and learning rate $\alpha$, poses a secondary optimization problem.  
In this work we used the PReLU activation function and fixed the batch size at 1200.  
The number of epochs trained for is problem dependent but is chosen manually to be large enough that training further would result in negligible decrease in error. 
Optimization of $\ell$, $H$, and $\alpha$ is performed through grid scans of parameter ranges to find hyperparameters that define neural networks that, when trained, achieve minimal loss. 
Generally, we find wide, shallow networks of a few layers and almost 1,000 nodes per layer perform best for the regression problems we are solving.

The open-source Python machine learning library \verb+PyTorch+ was used to implement the neural networks, which were trained on GPUs from the Perlmutter cluster at NERSC.

\begin{figure*}[ht]
    \centering
    \includegraphics[width=\linewidth]{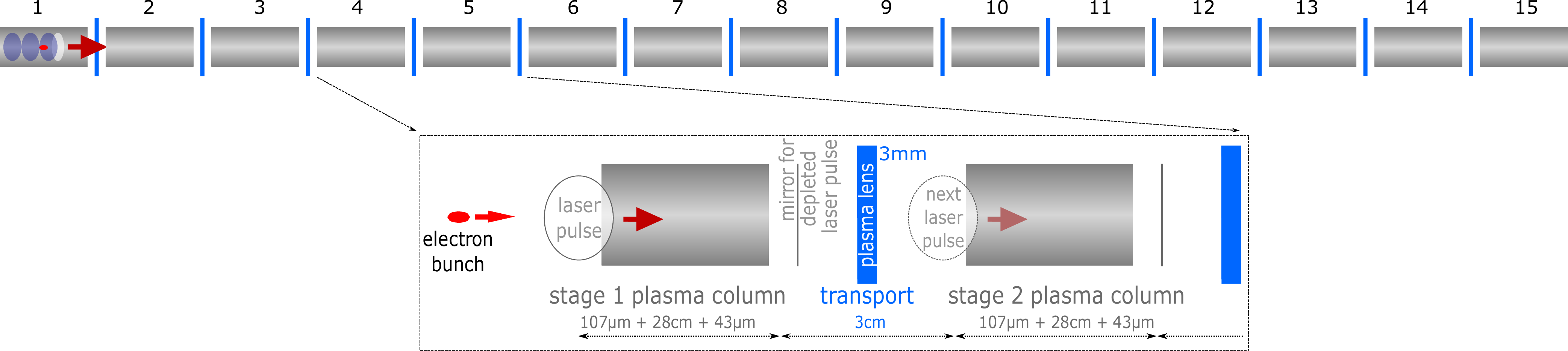}
    \caption{Schema of a chain of plasma accelerator stages.
    Grey rectangles mark the laser-plasma accelerator stages, blue lines denote the details of the transport gap (drift elements and focusing plasma lens) for the beam.
    Inset: detail schema of two plasma accelerator stages.
    With fast plasma column modeling, one can rapidly co-design significantly longer and more complex, apochromatic or non-linear transport gaps~\cite{LindstroemPRAB2021}.}
    \label{fig:15_stages}
\end{figure*}

The data to train the neural networks was generated with high-fidelity, GPU-accelerated \texttt{WarpX} computer simulations. 
The initial and final beam phase space coordinates of the particles in the particle bunch were collected, resulting in two $N\times n$ arrays corresponding to the initial and final particles' phase space coordinates.  
The data were cleaned by examining the final phase space images.  
Some particles were very clearly outliers and so were removed from the initial and final datasets.
In each coordinate direction, data were normalized to the mean and standard deviation.  
That is, for each phase space direction, the mean for that direction was subtracted from the data, which was then divided by the standard deviation for that direction.  
The computational particle data was then divided into training and testing subsets.  
70\% of the simulation particle data was used to update the neural network parameters in the training process while 30\% of the simulation particle data was reserved for testing.  
In the training loop, after the model weights were updated, the model was evaluated on the testing dataset.  
The error on the testing set was not used to update the weights, just to evaluate how well the model was able to generalize as it learned the training data.

\section{Benchmark: Staged Laser-Wakefield Simulations
\label{sec:benchmark}}
  
A challenging problem in particle accelerator design is the optimization of the accelerator beamline for best beam quality.
For example, the plasma-based accelerator community is working to demonstrate plasma-based accelerators that achieve collider-relevant energies and beam quality~\cite{Steinke2016,LindstroemPRAB2021}.
A simplified version of this problem involves repeated identical laser-plasma accelerator (LPA) elements joined with simple transport sections consisting of vacuum propagation, a small idealized focusing element, and further vacuum propagation, as depicted in Fig.~\ref{fig:15_stages}.
The LPA elements are optimized for energy gain and beam quality assuming the particle beam is properly matched. 
The focusing elements for the particle bunch in our example are plasma lenses~\cite{vanTilborg2015}.

In our idealized case the plasma lenses have two free parameters, focusing strength and position, which can be varied between each consecutive pair of LPA elements.
This results in $2(N_{\rm stage}-1)$ parameters to vary for $N_{\rm stage}$ LPA stages.  
The 3D simulations required to capture all the physics of laser-plasma particle acceleration can be expensive.  
While it is feasible to perform a few simulations for a given beamline, it would be prohibitively expensive to perform the hundreds of simulations required to fully explore the effects of varying the $2(N_{\rm stage}-1)$ lens parameters. 
In this section, we present the use of surrogate models for stages as an alternative to repeated direct simulation of the staged LPA beamline.

\begin{figure}[ht]
    \centering
    \includegraphics[width=\columnwidth]{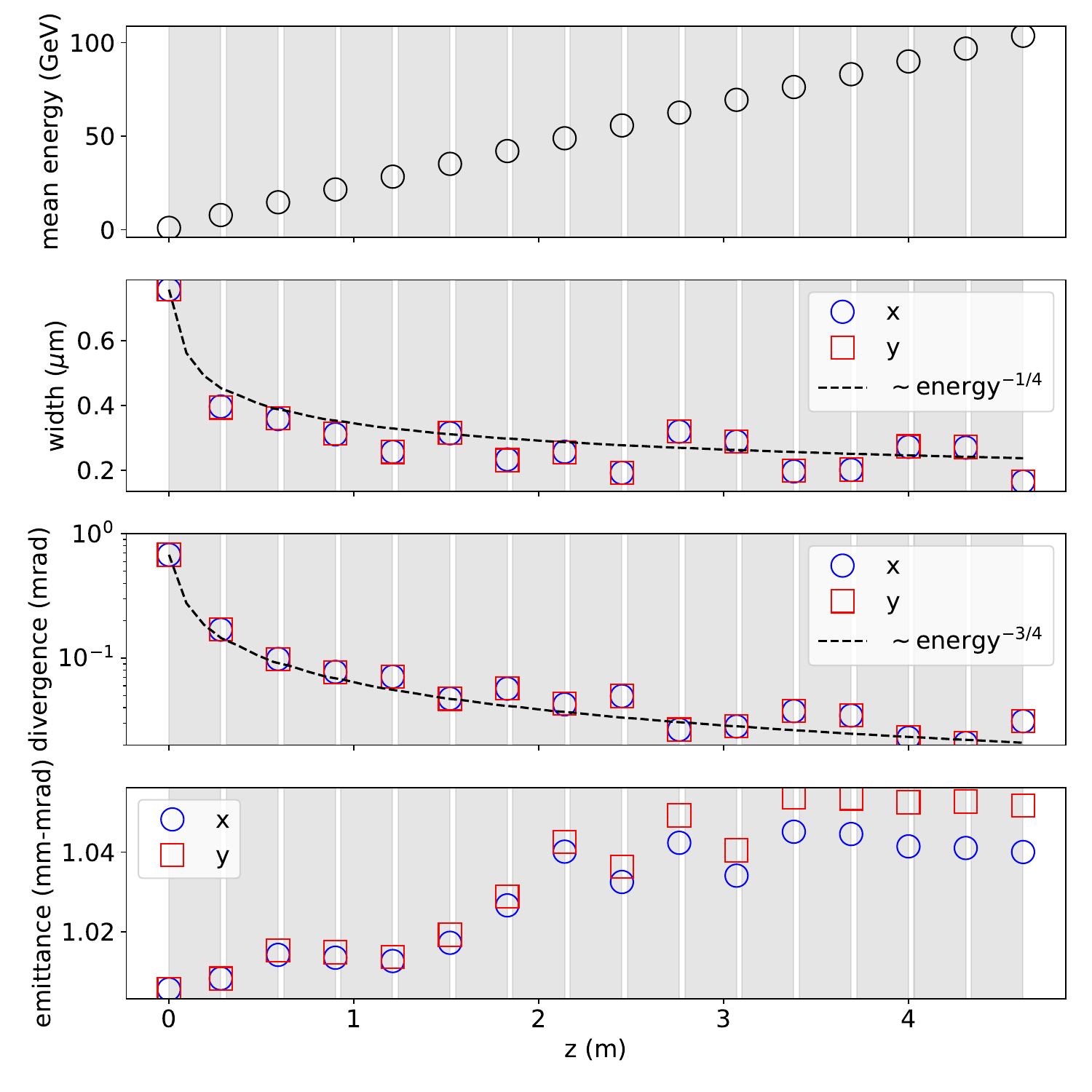}
    \caption{Electron beam evolution in 15 stages of chained laser-plasma accelerators as diagrammed in Fig.~\ref{fig:15_stages} from a \texttt{WarpX} simulation. 
    The top panel shows the energy gain. The second panel shows evolution of the beam width. 
    The third panel shows the beam divergence. 
    The bottom panel shows the emittance. 
    Longitudinal extent of the stages is indicated with pale grey shading.}
    \label{fig:energy_emittance}
\end{figure}

Consider an $N_{\rm stage}=15$ stage layout, with initial electron beam parameters as indicated in table~\ref{tab:training_beams} and laser-plasma parameters chosen for about 7\,GeV energy gain per stage and to control the electron beam transverse emittance.
We refer to a \verb+WarpX+ simulation of this layout throughout this paper as a baseline for measuring performance of the \verb+ImpactX++Surrogate simulation.
The \verb+WarpX+ simulation gives access to statistical beam moments, as shown in Fig.~\ref{fig:energy_emittance}, and to all the particle data.
The top panel of Fig.~\ref{fig:energy_emittance} shows the average energy of the electron beam as a function of propagation distance $z$. 
Only a few points are plotted -- the initial value and the energy at the end of each plasma stage. 
The electron beam gains about 7\,GeV of energy in each stage.

The next three panels in Fig.~\ref{fig:energy_emittance} show the evolution of statistical beam moments in the coordinate directions transverse to the axis of propagation.  
The second panel shows the evolution of the beam width $\sigma_{x}=\sqrt{\langle x^2\rangle}$ with propagation distance $z$.
Here $\langle q \rangle\equiv \sum_{i=1}^N w_i q_i / \sum_{i=1}^N w_i$ means to take the average over the beam distribution of quantity $q=x,p_x/p_z,y,p_y/p_z,\ldots$, weighted with the particle mass or weight $w_i$.
In a plasma accelerator stage, there is a specific beam width at which the expansion of the beam is balanced by the focusing forces in the wakefield.
Combining the definition for beam width with the definition of emittance and the focusing strength in the wakefield~\cite{LindstroemPRAB2021}, we find that the matched beam width varies with the inverse fourth root of beam energy $E$, i.e. $\sigma_x\sim E^{-1/4}$.
This ideal evolution of the beam width is indicated with the dashed black line.
Note that the beam width oscillates about the matched beam width -- this deviation from ideality suggests that the current transport parameters can be improved on.
The third panel shows the evolution of the beam divergence $\sigma_{\theta_x}=\sqrt{\langle {(p_{x}/p_z)}^2 \rangle}$. 
Note that all momenta $p_x$, $p_y$, and $p_z$ in this work are scaled by electron mass times the speed of light and so are dimensionless. 
As for the beam width, the matched beam divergence is indicated with the dashed black line. 
The matched divergence varies with the inverse cube of the fourth root of beam energy $E$, that is, $\sigma_{\theta_x}\sim E^{-3/4}$~\cite{LindstroemPRAB2021}.

\begin{table*}[ht]
    \centering
            \begin{tabular}{|c|c|c|c|c|c|c|}
            \hline 
                  & $\sigma_x=\sigma_y$ ($\mu$m) & $\sigma_z$ ($\mu$m) & $\sigma_{px}$ & $\epsilon_{nx}$ (mm-mrad) & energy (GeV) & relative energy spread (\%)\\
                 \hline
                 staged simulation & $0.75$ & $0.1$ & 1.33 & 1 & 1 & 0 \\
                 \hline
                 stage $i$ & $2$ &2 &8 &16 & $\approx 1+7i$&10 \\
                 \hline
            \end{tabular}
    \caption{Initial beam parameters for the 15-stage simulation and for the training simulations.  
    The energies listed for the training stages are approximate as these were set more precisely to the exact energies of the beam observed in the staged \texttt{WarpX} simulation.}
    \label{tab:training_beams}
\end{table*}

The final panel shows the evolution of the normalized transverse beam emittance $\epsilon_{x}=\sqrt{\langle x^2\rangle\langle p_x^2\rangle - \langle x p_x\rangle^2}$. 
Note that the y quantities are similarly defined. 
For use in a collider, emittance needs to be preserved.
The 5\% growth in emittance over 15 stages is still too large for collider applications and now needs to be reduced.


The \texttt{WarpX} reference simulation in this study took about 1,316 GPU-hours using 64 nodes (256 Nvidia A100 GPUs) of the Perlmutter machine at NERSC, which we consider to be moderately well resolved.


Beam transport between stages is governed by the strength of the focusing optics. 
The goal of the surrogate optimization is to maintain energy growth while suppressing emittance growth.  
In this case the focusing optics are idealized lenses with linear focusing forces.  
The lenses for this \texttt{WarpX} simulation are centered between the plasma stages and the lens strengths are determined by solving an analytic expression for lens strength using the first order electron beam average energy $\gamma_{ave}=\langle \gamma \rangle$ and the second-order beam moments beam width $\sigma_x$, divergence $\sigma_{\theta_x}$, and position-velocity correlation $\sigma_{x\theta_x}=\langle x p_x/p_z\rangle$. 
This technique for setting the lens strengths was derived with a collaborator and first demonstrated in Ref.~\cite{Pousa2023}.
These beam moments are evaluated in the simulation after each plasma stage, in the gap between the plasma and the lens.

The LPA stages in this example can be understood as nonlinear phase space maps that require PIC simulations to compute, so this example is a natural candidate for hybrid surrogate/conventional simulations.  
To obtain the surrogate models, we train neural networks to learn the phase space map $f:\mathbb{R}^6\to\mathbb{R}^6$ from initial to final beam phase space through the stages.  

\subsection{Training simulation for neural network surrogates}
The training data is generated from a single, high-fidelity \texttt{WarpX} HPC simulation.
The simulation models 15 particle bunches of different initial energy going simultaneously through a single laser plasma stage.
The different bunches are set to very low charge, so they can pass through the accelerator simultaneously.
This is a common particle-tracking approach that neglects space charge effects.
The training particle bunches each consist of $N=10^6$ particles.
They are centered at the initial energy of the corresponding bunch from the staged simulation; that is, the first bunch starts at the initial simulation energy of 1\,GeV, the second bunch starts at the energy of the bunch from the staged simulation going into the second stage, etc.
The training bunches are chosen to have a relative energy spread of 10\%, bunch duration 6.67\,fs, widths $\sigma_x=\sigma_y=2\ \mu{\rm m}$ and momentum widths $\sigma_{px}=\sigma_{py}=8$.  Note that in the figures we plot the beam divergence $\sigma_{\theta_x}=\sqrt{\langle {(p_{x}/p_z)}^2 \rangle}$.

\begin{figure}[ht]
    \centering
    \includegraphics[width=\linewidth]{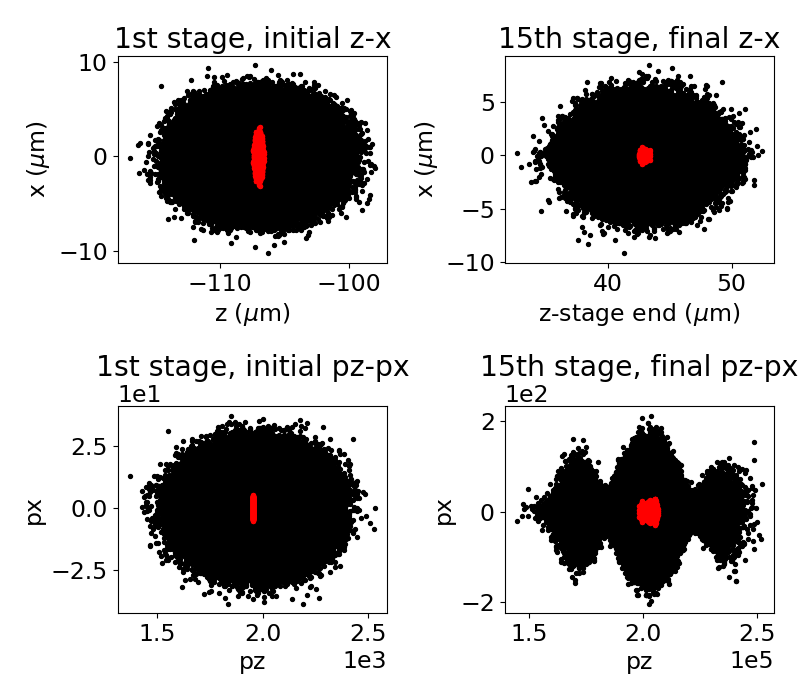}
    \caption{Comparing the training and staged beams. 
    Particles from the training simulation are  black $\bullet$, from the staged simulation are {\color{red}red $\bullet$}. 
    Top-left: $z$-$x$ staged and training phase spaces initially, bottom-left: $p_z$-$p_x$ staged and training phase-spaces initially, top-right: $z$-$x$ staged and training phase space at the end of stage 15, bottom-right: $p_z$-$p_x$ staged and training phase spaces at the end of stage 15.
    }
    \label{fig:training_v_simulation}
\end{figure}

The training bunches are chosen to be centered on the expected bunch parameters but several times larger in each phase space direction, as seen in Fig.~\ref{fig:training_v_simulation} to allow the surrogates to learn a broader region in phase space and generalize to phase space distributions beyond what is in the staged simulation to facilitate exploration of the lens parameters in the optimization. 
We introduce $\xi=z-ct$ for the relative longitudinal position.
The larger the region of phase space that the models learn, the more work it is to train the models.  
Deciding the region of phase space to cover with the training bunches is a trade-off between minimizing computational effort and maximizing generalizability of the models. 
The region chosen was found to be a reasonable compromise between model accuracy and training time. 

The particle phase-space positions of the training bunches are initially sampled from a Gaussian distribution, as are the bunch particle positions in the staged simulation.  
The use of a Gaussian distribution in generating training data is primarily a matter of convenience, as this is already implemented in \texttt{WarpX}.  
The Gaussian distribution is reasonable for the purposes of this example, however, as most data is centered near the core of the training bunch, which is the region where the staged simulation begins and where the bunch at subsequent stages is expected to be. 
For example, in Fig.~\ref{fig:training_v_simulation}, the left two panels show the region of phase space initially sampled by the staged and training beams; we see that the staged bunch is initially a small sliver of the training beam.  
The right panels show the relative overlap of the staged beam at the end of the final  stage with the training bunch for the second final.  
We see that, compared to the initial conditions, the staged bunch is larger relative to the training beam, but still occupies only a fraction of the phase-space area of the  training bunch. 
The training bunch parameters are described in table~\ref{tab:training_beams} and a complete description of the training simulation can be found in the supplementary material~\cite{Data}.
The training simulation took about 404 GPU-hours on 64 nodes (256 GPUs) of Perlmutter.

\subsection{Training neural network surrogates}
    
    The training data is collected from the training simulation and stored as initial and final $N\times 6$ arrays as discussed in section \ref{sec:neural_network_surrogates}. 
    These input and output arrays allow us to perform supervised learning. 
    In this work we used fully connected feed-forward neural networks with PReLU activation functions.
    The neural networks for the first three stages have 5 hidden layers and 900 nodes per layer.
    The neural networks for the last 12 stages have 3 hidden layers and 700 nodes per layer.  
    Hyperparameter tuning was done by grid search for the surrogates for the first 5 stages, after which it was observed that similar models yield similar accuracy.
    For these parameters, training on one A100 GPU at Perlmutter took about 2.2 hours per stage for the first 3 stages and about 2 hours per stage for the last 12 stages.

\begin{figure}[ht]
            \includegraphics[width=\columnwidth]{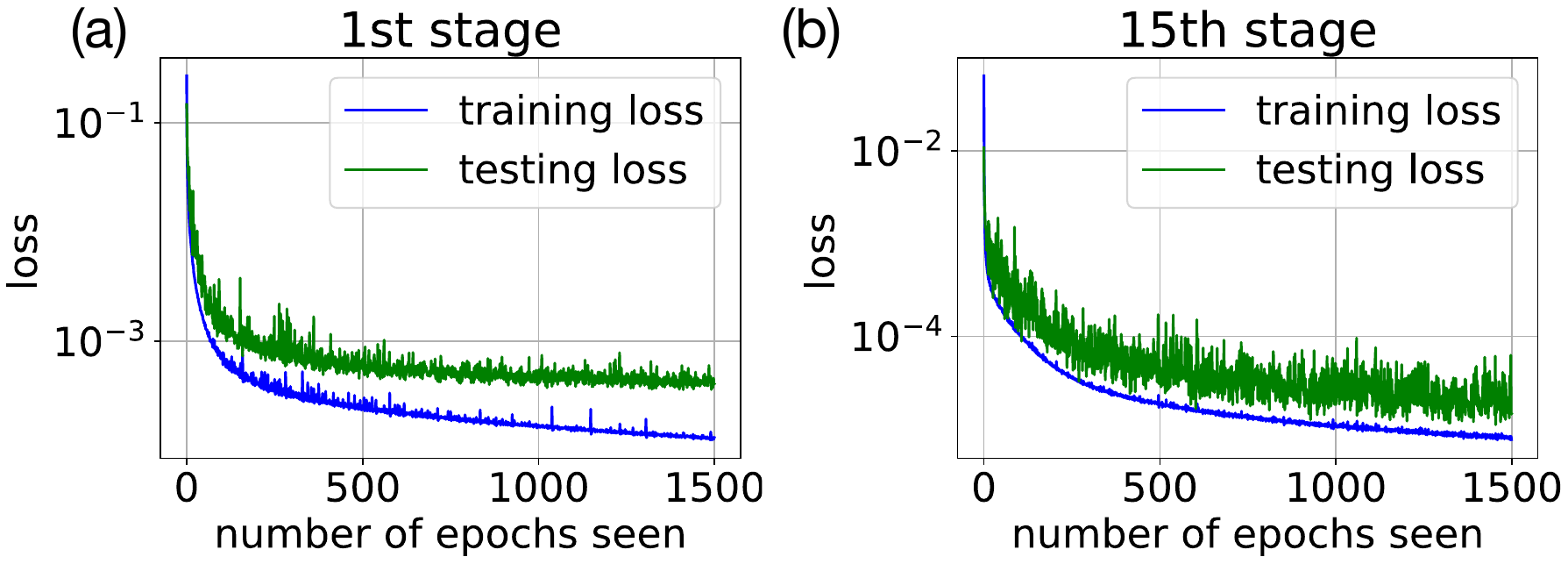}
            \caption{\label{fig:train_test_loss}This figure shows training (in {\color{blue}blue}) and testing (in {\color{green}green}) loss from the training of the neural networks for certain stages.  
            In (a) are the training and testing losses from stage 1 and (b) are the loss curves from stage 15.}
        \end{figure}

    We used the MSE loss function. 
    While we experimented with other loss functions, including a physics-inspired custom loss function that penalizes second order terms $x_{ij}x_{ik}$, we did not see significant improvement over the MSE loss. 
    For training we found a learning rate of $1\times10^{-4}$ works best. Training and testing loss curves can be seen in figure \ref{fig:train_test_loss} for stages 1 and 15. 
    The curves for stages 2 through 14 are similar to the stage 15 curve. 
    While the training loss has visible slope at the 1,500 epochs, indicating that further training would decrease the training loss and lead to more accurate predictions on the training data, the test loss has less slope and is nearly flat for stage 1.  
    This indicates that the model is not getting any better at interpolating to unseen data and further training will not make the model more accurate generally. 
    Note that the achievable loss decreases with increasing stage number.  
    This corresponds with the more rigid particle dynamics at higher average bunch energies -- 
    at higher energies, particles are more relativistic, meaning less transverse motion in the same longitudinal distance.   
    Rigid dynamics means the neural network has an easier time learning the map to the final particle phase space at later stages than at earlier stages.

\begin{figure}[ht]
        \centering
        \includegraphics[width=\linewidth]{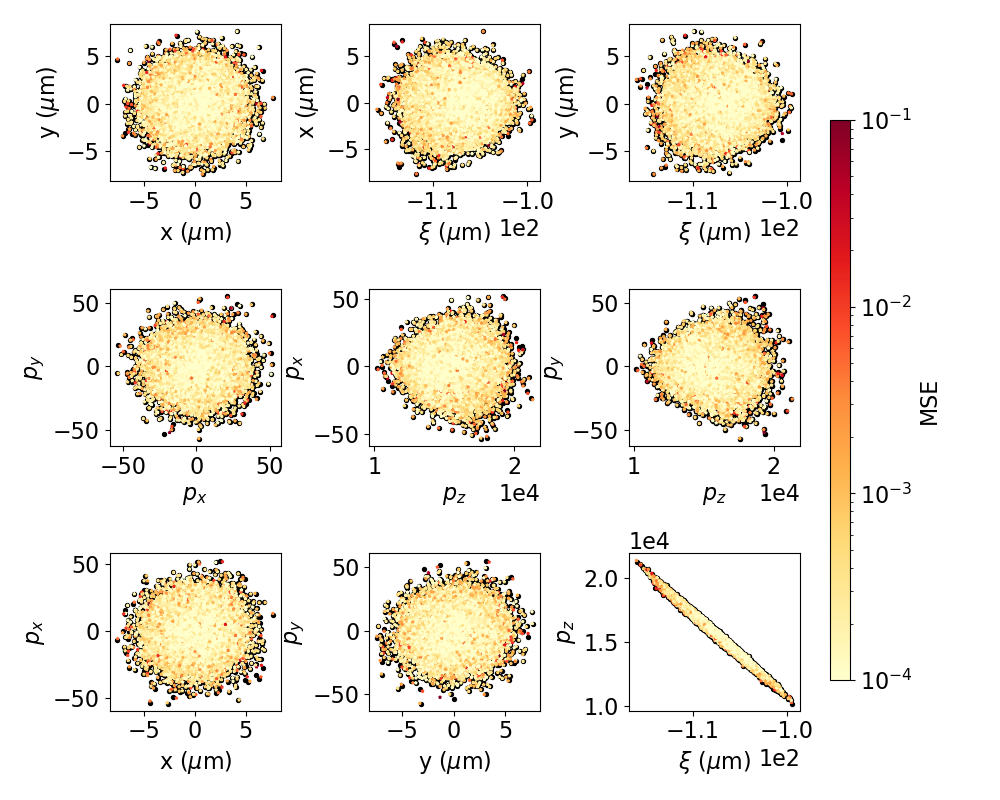}
        \caption{Training stage 1: Comparison of bunch phase spaces from the bunch for stage 1 in the \texttt{WarpX} training simulation ({black$\bullet$}) and the predicted final phase space of this bunch as calculated by the neural network to be used as a surrogate for stage 1 ({\color{yellow}yellow$\bullet$}-{\color{red}red$\bullet$}, colored by MSE loss). 
        Top row: spatial projections, middle row: momentum projections, bottom row: phase space projections.}
        \label{fig:visual_beam_comparison_stage_0}
    \end{figure}

    We gain some insight into what the model is trying to learn from visually comparing the bunch phase space from the training data with the predicted phase space.  
    Fig.~\ref{fig:visual_beam_comparison_stage_0} shows the phase space of the beam used to train for stage 1 in black and the predicted phase space of the bunch in color.  
    Each particle of the predicted phase space is colored by its mean-squared error after normalization.  
    To make sure the simulation bunch is visible behind the predicted bunch, the black dots are also  larger, leading to a black border around the predicted bunch that should not be mistaken for error in the prediction. 
    The top row shows the $x$-$y$, $z$-$x$, and x-y spatial projections.  
    In the $x$-$y$ projection, the bunch is symmetric and error increases radially from the center.
    Since our training data is sampled from a Gaussian distribution, we attribute the lower error in the beam center to there being more available training data for this region of phase space.
    The $x$-$z$ and $y$-$z$ projections are similar, reflecting again the azimuthal symmetry.  
    The middle row shows the $p_x$-$p_y$, $p_z$-$p_x$, and $p_z$-$p_y$ momentum projections.  
    The spatial-velocity coupling in each direction leads to similar patterns in these momentum projections as in the spatial projections of the top row.
    The bottom row shows the $x$-$p_x$, $y$-$p_y$, and $z$-$p_z$ phase spaces.  
    In $x$-$p_x$ and $y$-$p_y$ we see the bunch is elliptical and again showing the increase in error with radial distance attributed to the decreasing information density with radial distance.  
    The $z$-$p_z$ phase space shows the strong coupling between longitudinal position and energy characteristic of the plasma-based acceleration and the micron-scale accelerating structure. 
    The tight agreement between the phase spaces overall reflects how well the neural network learns the particle phase space. 
    As seen in Fig.~\ref{fig:training_v_simulation}, the phase space region of most interest is in the center of the training bunch, where the model accuracy is highest.
    
    The phase spaces for the other training bunches are similar in shape, features, and overall agreement of the neural network prediction with the simulation.  

\subsection{Incorporating surrogates into \texttt{ImpactX}}
The surrogate models are seamlessly incorporated into an \verb+ImpactX+ beamline simulation as discussed in Sec.~\ref{sec:impactx_and_surrogates}.
Inference of the neural network surrogate models for 10,000,000 particles, as is used in the staged simulation, takes about 632\,ms/stage on an Nvidia A100 80 GB SMX and scales roughly linearly with the number of particles and number of stages, giving an estimated inference cost of 63.2\,ns/particle/stage.

More precisely, the surrogate push has a time of about 1.12\,s for the first three neural networks and 494\,ms for the last twelve networks.  
Recall that the first three networks have two more hidden layers and each hidden layer is about 25\% larger.
For the first three stages, about 1.108\,s are spent in inference, and the rest of the time is spent in preparing the data for/after inference.
That is, 99\% percent of the surrogate push is spent running GPU kernels to apply the neural network, showing the efficiency of the all-GPU data access and preparation.
For the last 12 stages, 483 of the 494\,ms of the surrogate push are spent in neural network inference, or 98\% of the surrogate push is spent in inference and 2\% is spent on data preparation.
Data preparation are applying/reversing normalization factors and array-to-struct transformations (arrays to/from tensors), all on GPU.
There is no movement of the beam particles from GPU to CPU and all GPU operations are executed in CUDA streams in the typical, asynchronous manner.

\begin{figure}[ht]
    \centering
    \includegraphics[width=\columnwidth]{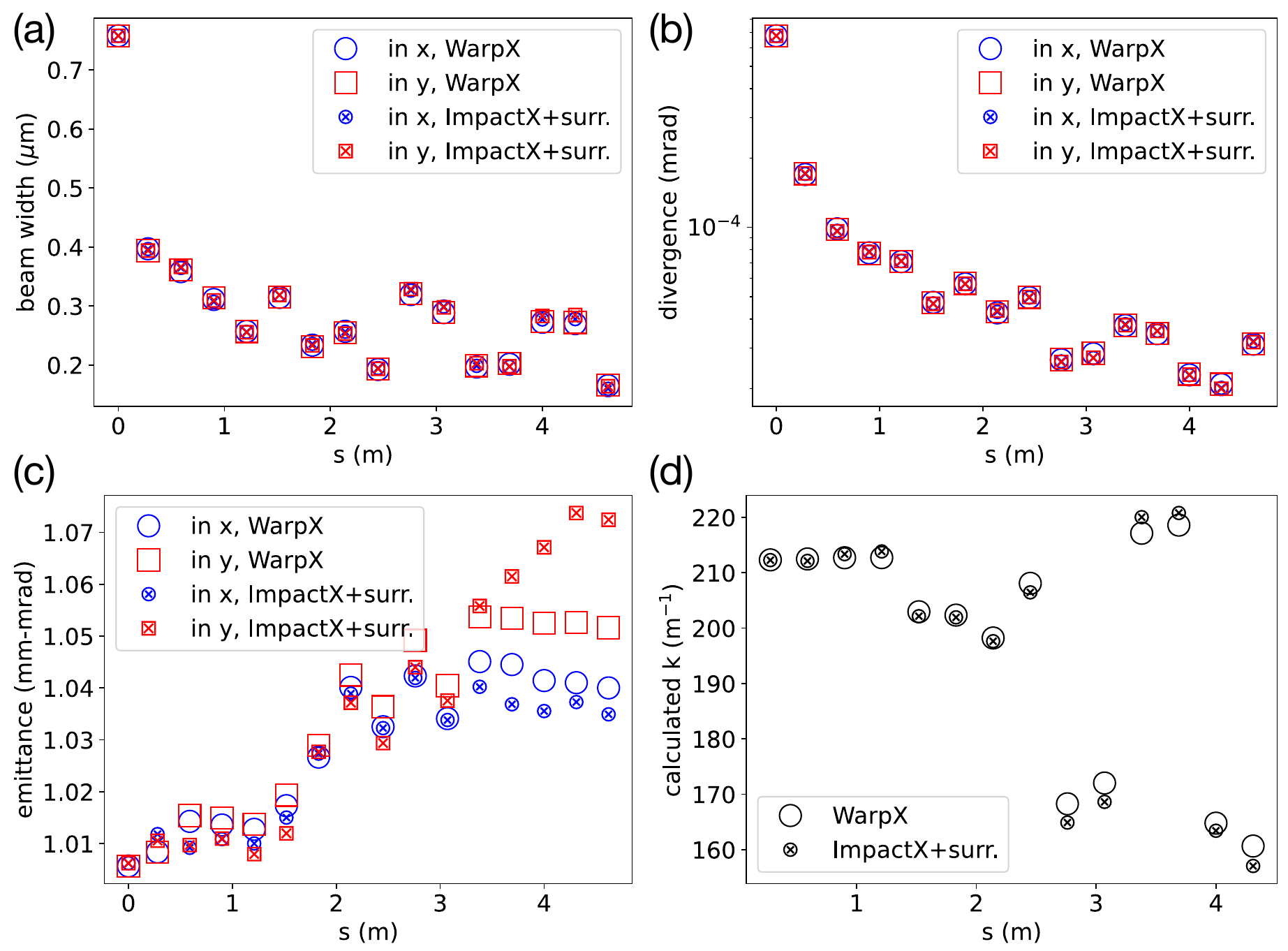}
    \caption{Comparison of bunch moment evolution in \texttt{WarpX} simulation with \texttt{ImpactX} simulation using surrogate models. (a) beam width $\sigma_{x,y}$, (b) beam divergence $\sigma_{\theta_{x,y}}$, (c) beam emittance $\epsilon_{x,y}$, (d) calculated lens strengths}
    \label{fig:compare_warpx_impactx_surrogate}
\end{figure}

Fig.~\ref{fig:compare_warpx_impactx_surrogate} shows the evolution of bunch moments in the \verb+ImpactX+ + surrogate and \verb+WarpX+ simulations.  
Fig.~\ref{fig:compare_warpx_impactx_surrogate}(a) shows the evolution of bunch spot size over propagation distance, plotted initially and at the end of each stage. 
Fig.~\ref{fig:compare_warpx_impactx_surrogate}(b) shows the evolution of the bunch divergence, which  oscillates similarly to bunch size.   Fig.~\ref{fig:compare_warpx_impactx_surrogate}(c) shows the evolution of the bunch emittance, which grows as a result of the beam mismatch into each stage and beam oscillation in phase space as shown in (b) and (c).

The x- and y-emittances predicted by the \verb+ImpactX+ + surrogate simulation agree remarkably well with the reference \verb+WarpX+ emittances through the first 10 stages.
In the final 5 stages, the x-emittance between the two simulations agrees closely but the y-emittance from the \verb+ImpactX+ + surrogate simulation begins to differ from the reference simulation by a few percent.
Moreover, it shows qualitatively different behavior by increasing whereas the y-emittance of the reference simulation saturates.
We do not know the origin of this discrepancy both from anticipated x-y symmetry and from the reference simulation, but it is only on the order of a few percent and there are many possible sources of random error:
Significant among these is that each stage is modeled with a neural network that has randomly initialized parameters that are then refined by stochastic gradient descent.
There is statistical noise inherent in the simulations and so in the training data that results from randomized beam initialization.

Ultimately, the agreement between the surrogate and reference simulation is very good through the first 10 stages and even to within a few percent through the full 15 stages.
Fig.~\ref{fig:compare_warpx_impactx_surrogate}(d) shows the lens strengths used in each simulation.  
Recall that the lens strengths are calculated in the course of the simulation, using the bunch parameters as the bunch is about to enter the lens.  
The lens strengths are a high-level quantity requiring accurate calculation of several sensitive second-order bunch moments throughout the simulation, which in turn requires very accurate surrogate models for the LPA stages.  
Hence the close agreement of the two methods shown in Fig.~\ref{fig:compare_warpx_impactx_surrogate}(d) is a strong indication of the potential of the surrogate models to capture the dynamics of the full 3D PIC simulation and the potential of the \verb+ImpactX+ + surrogate simulations for accurate exploration of parameter space. 
We see that the \verb+ImpactX+ + simulation with LPA surrogates is able to predict the bunch width, divergence, and emittance evolution of the 3D \texttt{WarpX} simulation to within a few percent. 
The mean bunch energies are not plotted here but are similarly well-predicted by the \verb+ImpactX+ + surrogate simulation.

\begin{figure}[ht]
    \centering
    \includegraphics[width=\columnwidth]{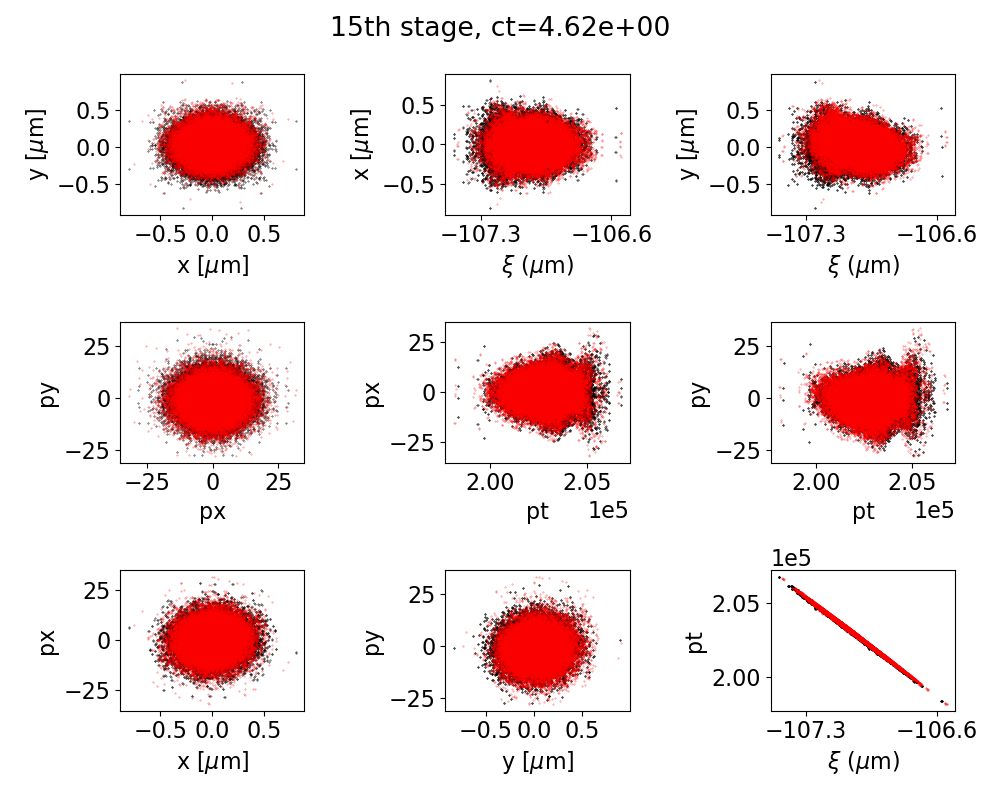}
    \caption{Benchmark: Comparison of beam phases from \texttt{WarpX} (black$\bullet$) and \texttt{ImpactX}+surrogate ({\color{red}red$\bullet$}) simulations at stage 15. 
    Top row: spatial projections, middle row: momentum projections, bottom row: phase space projections.}
    \label{fig:compare_phase_warpx_ixs_stage_8}
\end{figure}

A visual inspection of the bunch phase space at the end of stage 15 is shown in Fig.~\ref{fig:compare_phase_warpx_ixs_stage_8}.
In contrast with Fig.~\ref{fig:visual_beam_comparison_stage_0}, the most interesting figure is for stage 15 rather than for stage 1.  
Here the interest is not the performance of the model on a given stage but how all the models perform in aggregate. 
The bunch particles from the \verb+WarpX+ staged simulation are shown in black and the bunch particles of the \verb+ImpactX+ + surrogate simulation are in red.
The top row shows the $x$-$y$, $z$-$x$, and $x$-$y$ spatial projections.  
As in Fig.~\ref{fig:visual_beam_comparison_stage_0}, in $x$-$y$ the beam is symmetric.  
The $x$-$z$ and $x$-$y$ phase spaces are similar, reflecting again the azimuthal symmetry.  
The middle row shows the $p_x$-$p_y$, $p_z$-$p_x$, and $p_z$-$p_y$ momentum projections.  
The spatial-velocity coupling in each direction leads to similar patterns in these momentum projections as in the spatial projections of the top row.
The bottom row shows the $x$-$p_x$, $y$-$p_y$, and $z$-$p_z$ phase spaces.  
In $x$-$p_x$ and $y$-$p_y$ we see the beam is experiencing some deviation from its initial elliptical shape, a result of imperfect matching and beam dynamics through the accelerator.  
The spatial energy correlation seen in the z-energy phase space is characteristic of laser-plasma acceleration and the microscopic acceleration structure.
The \texttt{ImpactX}+surrogate model reproduces the phase space projections of the \texttt{WarpX} simulation accurately, even capturing the small beam mismatching and evolution that are occurring in the 15 stage \texttt{WarpX} simulation.

The phase spaces at the end of earlier stages are similar in shape and features, with similar and even better agreement between the particles of the \texttt{WarpX} and \texttt{ImpactX}+surrogate simulations.

\subsection{Optimization Using Conventional-Surrogate Simulations \label{sec:optimization}}
The \verb+ImpactX++surrogate simulation ran quickly, facilitating the exploration of large regions of parameter space and even numerical optimization. 
For example, using the simplex or Nelder-Mead optimization \cite{Gao2012} for scalar optimization in the \texttt{SciPy} optimization package~\cite{scipy} in Python, the optimal lens parameters were determined for minimizing the emittance in $x$ in the \verb+ImpactX++surrogate simulation. 
This was performed as a series of optimizations, minimizing the emittance in $x$ after stage 1 by varying the plasma lens position and strength between stages 1 and 2; then minimizing the emittance in $x$ after stage 2 by varying the plasma lens position and strength between stages 2 and 3; and so on until finally minimizing the emittance in $x$ after stage 15 by varying the plasma lens position and strength between stages 14 and 15.
After optimizing the \texttt{ImpactX}+surrogate simulation, the resulting best lens positions and strengths were transferred back to a \texttt{WarpX} simulation for verification in full-fidelity.
The parameters of the \texttt{WarpX} simulation were identical to the reference simulation other than the new lens parameters.

\begin{figure}[b]
    \centering
    \includegraphics[width=\columnwidth]{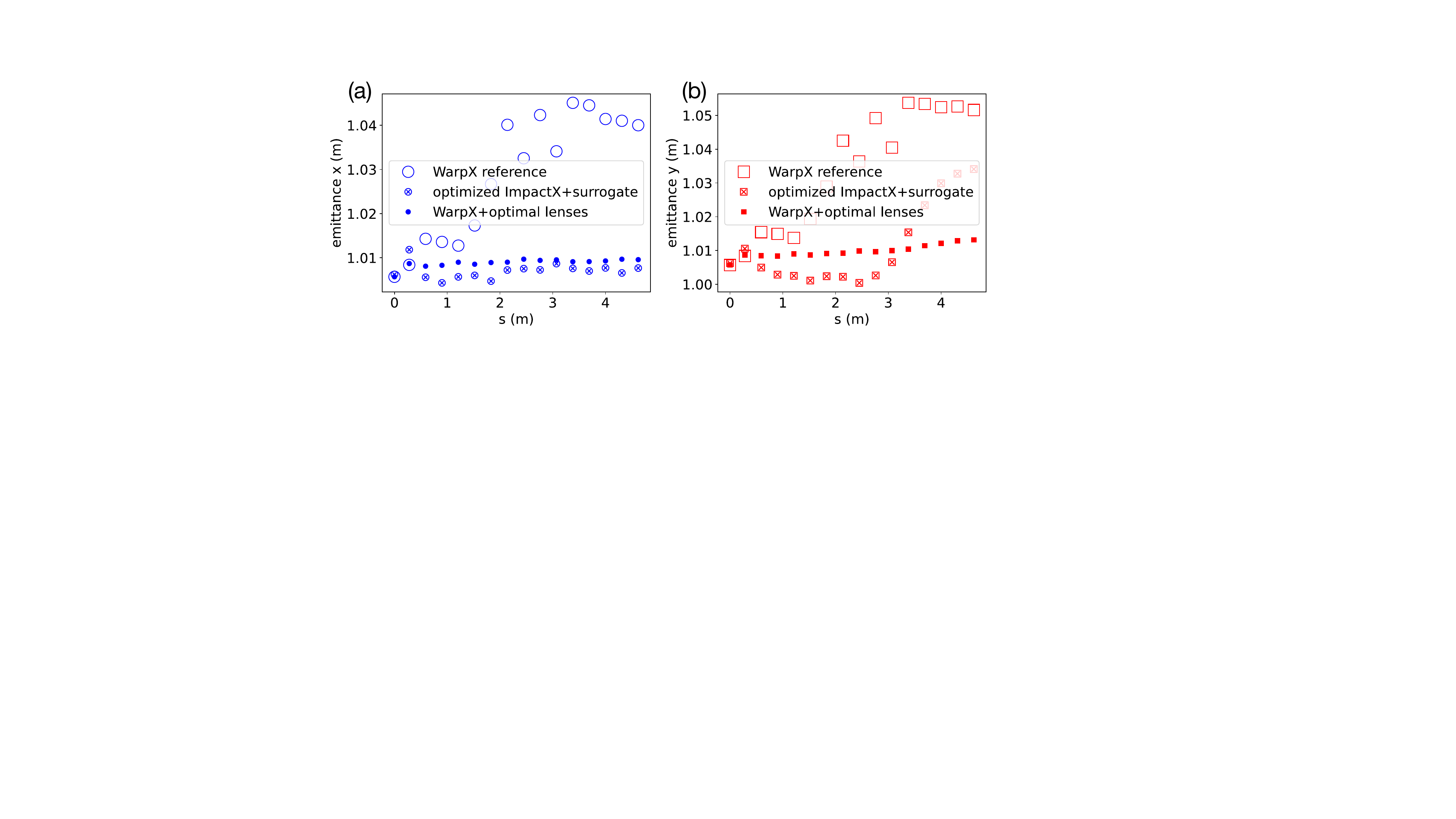}
    \caption{Optimization: Comparison of emittances in (a) $x$ and (b) $y$, between the \texttt{WarpX} reference simulation, the optimized \texttt{ImpactX}+surrogate simulation, and the \texttt{WarpX} simulation using the found optimal lens parameters.
    Lower emittance values are better.}
    \label{fig:optimized_emittance}
\end{figure}

The resulting emittances can be seen in Fig.~\ref{fig:optimized_emittance}.  
Figures~\ref{fig:optimized_emittance} (a) and (b) show the normalized emittances in $x$ and $y$, respectively.
The hollow blue circles indicate the emittance of the bunch through the \verb+WarpX+ reference simulation, the blue circles with crosses indicate the emittance of the bunch through the optimized \verb+ImpactX++surrogate simulation, and the solid blue circles show the emittance of the bunch through the \verb+WarpX+ simulation with optimal lenses.
The same fill pattern of hollow, crossed, and solid for the red squares indicates the emittances in $y$.

The optimization resulted in emittances that improve on the previous \verb+ImpactX++surrogate simulation and that stay constant after the first stage.
The surrogate models are accurate enough that the optimized \verb+ImpactX++surrogate parameters preserve emittance in the full-fidelity \verb+WarpX+ simulation as well.
Notably, although the optimization was for emittance in the $x$ direction, the emittance in the $y$ direction is also improved.
While the emittance in $y$ does ultimately increase over the last several stages in the optimized \verb+ImpactX++surrogate simulation, emittance growth saturates at less than what is observed without the optimization.
In the \verb+WarpX++ simulation with the optimized lens parameters, emittance in $y$ is also almost constant.
There is a slight increase, but emittance is much better preserved compared to the \verb+WarpX+ reference simulation.

\begin{figure}
    \centering
    \includegraphics[width=\columnwidth]{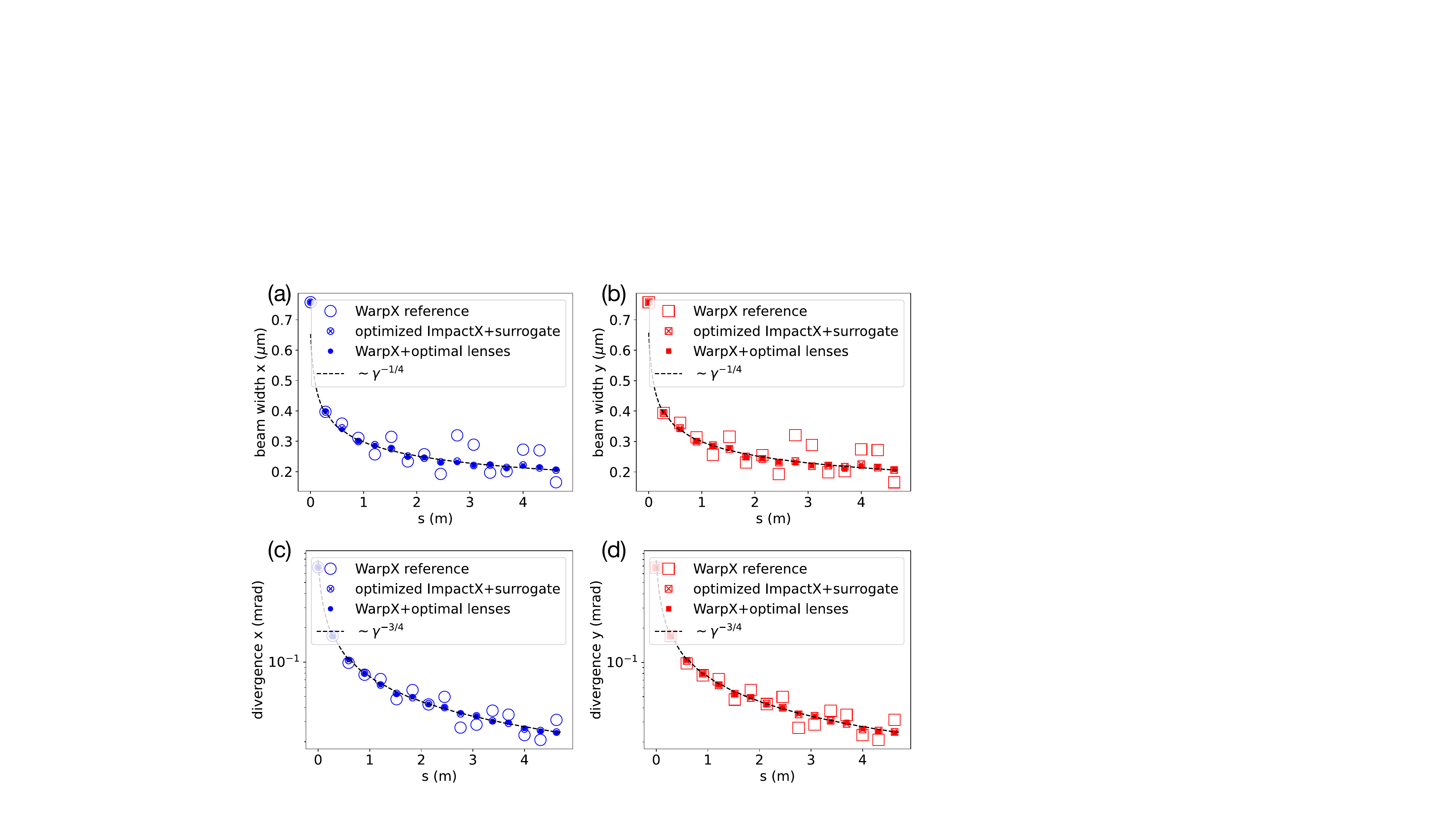}
    \caption{Optimization: Comparison of beam widths in (a) $x$ and (b) $y$ and divergences in (c) $x$ and (d) $y$, from the \texttt{WarpX} reference, the optimized \texttt{ImpactX}+surrogate simulations, and the \texttt{WarpX} simulation using the found optimal lens parameters.
    Note how the optimized values oscillate less around the dashed black line, which indicates the ideal beam moment evolution given perfect matching of the beam into each stage.}
    \label{fig:optimized_width}
\end{figure}

The resulting beam size and divergence can be seen in Fig.~\ref{fig:optimized_width}.
Figures \ref{fig:optimized_width} (a) and (b) show the divergence in x and y, respectively and Figures \ref{fig:optimized_width} (c) and (d) show the beam width in x and y, respectively.  
Recall, as was stated earlier, that if the beam is perfectly matched to each accelerator stage then the spot size should vary with the inverse fourth root of beam energy $E$ indicated with the dashed black line, $\sigma_x\sim E^{-1/4}$.
Note that the optimized beam width more closely follows the desired trend.
The optimization has improved on the analytical transport parameters used in the reference simulation.
This sort of optimization would be prohibitively expensive with PIC simulations.
Similarly, the divergence should vary with the inverse cube of the fourth root of beam energy $E$, i.e. $\sigma_{\theta_x}\sim E^{-3/4},$ indicated by the dashed black line, and the  divergence of the optimized simulation more closely follows the desired trend.


\section{Conclusions and future work}
There are large amounts of data from high-fidelity simulations.  
In this work, we utilize this data to develop data-driven models and workflows to complement existing HPC tools.
We developed a hybrid conventional-surrogate simulation that has shown three advantages: it is fast, reasonably accurate, and provides a flexible infrastructure for exploring problem structures and parameters. 

The simple neural network surrogate models in this work took each about 2 GPU-hours to train and the entire hybrid conventional-surrogate simulation takes 10 GPU-seconds (or a couple of CPU-minutes) to evaluate, whereas a single well-resolved PIC simulation takes >1,000 GPU-hours to run.
The hybrid conventional-surrogate simulation agreed with the high fidelity simulation to within a few percent in calculating the sensitive second-order quantities, even after accumulated errors through 15 stages.  
The derived quantities such as the predicted lens parameters for beam transport were similarly accurate to within a few percent.

The flexible hybrid conventional-surrogate infrastructure permits the easy rearrangement of accelerator elements or variation of parameters.
For example, we took advantage of the fast evaluation of the hybrid conventional-surrogate simulation to optimize for beam quality over 28 parameters in the staged LPA example.
The optimization significantly improved final bunch parameters.
Additionally, other interesting workflows can be envisaged including large parameter scans for fast plasma and transport design iterations, or sensitivity analysis.

The surrogate models in this work are simple and there is good reason to believe that more sophisticated neural network structures could be designed to better capture moments of the bunch distribution, reflect the symmetries of our problems, and/or encapsulate more of the relevant physics of the problems we study. 

While tracking simulations are generally valuable for accelerator modeling, surrogate modeling of the \textit{collective} effects of highly-charged beams poses an opportunity for future research.
The complexity of modeling collective effects changes fundamentally how the problem of predicting beam dynamics is approached – pure single-particle tracking is not effective because the particles interact with each other and the laser-plasma stage.
One step towards learning collective effects could be including beam moments in addition to the phase-space coordinates of each particle.

We would like to highlight that even without space charge effects, our ML models are already desirable and helpful to design detailed beam transport and optimize start-to-end accelerator lattices with chromatic effects, as we are actively studying in staged laser-wakefield acceleration for future high-energy physics colliders. 
With our model, we can test and design new complex (e.g., chromatic) beam transport via \texttt{ImpactX} for many beam phase space configurations without having to rerun the costly \texttt{WarpX} high-fidelity simulation.
We believe that the herein presented HPC software improvements and training workflows are an ideal starting point for such complex research topics.

In our reference study of the staged laser-plasma acceleration in traditional simulations with \verb+WarpX+, we benefited from being able to compare to an analytic model~\cite{Pousa2023} for the bunch focusing and transport that provided a good benchmark for verification of the hybrid conventional-surrogate simulation.  
With the hybrid simulation verified, we can extend studies of hybrid beamlines to more complex beam transport for which there is no such good analytic model.  
This hybrid conventional-surrogate model is faster to evaluate than traditional reduced-physics models.
Trained on high-fidelity simulation data, surrogate elements do not add simplifications on the physics at play, while providing similar ability to predict beyond what is known analytically.

The hybrid conventional-surrogate simulation concept for particle accelerator simulations provides a powerful, data-driven modeling capability to complement the current full-scale and reduced-order computational tools.

\begin{acks}
We acknowledge the \texttt{ImpactX}, \texttt{WarpX} and \texttt{AMReX} open source communities for their invaluable contributions.
In addition, we acknowledge and thank Alexander Sinn and Thierry Antoun for their contributions to particle data structures~\cite{AMReXIJHPCA2024}.
This work was supported by the Laboratory Directed Research and Development Program of Lawrence Berkeley National Laboratory under U.S. Department of Energy Contract No. DE-AC02-05CH11231. 
This material is based upon work supported by the U.S. Department of Energy, Office of Science, Office of High Energy Physics, General Accelerator R\&D (GARD), under contract number DE-AC02-05CH11231.
This material is based upon work supported by the CAMPA collaboration, a project of the U.S. Department of Energy, Office of Science, Office of Advanced Scientific Computing Research and Office of High Energy Physics, Scientific Discovery through Advanced Computing (SciDAC) program.
This research was supported by the Exascale Computing Project (17-SC-20-SC), a joint project of the U.S. Department of Energy's Office of Science and National Nuclear Security Administration, responsible for delivering a capable exascale ecosystem, including software, applications, and hardware technology, to support the nation's exascale computing imperative.
This research used resources of the National Energy Research Scientific Computing Center, a DOE Office of Science User Facility supported by the Office of Science of the U.S. Department of Energy under Contract No. DE-AC02-05CH11231 using NERSC award HEP-ERCAP0023719.
\end{acks}

\section*{Data Availability Statement}
The data that support the findings of this study are openly available on Zenodo.org at \url{http://doi.org/10.5281/zenodo.11063781}.
The open source simulations that were used for the findings of this study are available at \url{https://github.com/ECP-WarpX/impactx} and \url{https://github.com/ECP-WarpX/WarpX}.

\appendix

\section{Appendix: Optimal lens parameters}
\begin{figure}[b]
    \centering
    \includegraphics[width=\columnwidth]{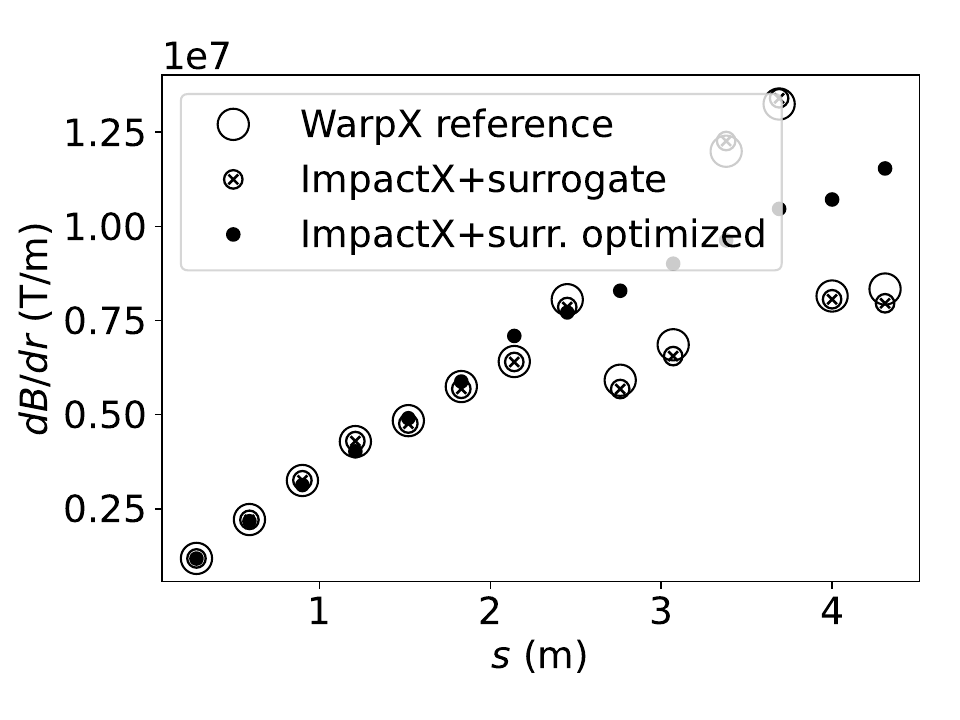}
    \caption{Lens strengths: Comparison of the lens strengths used in the WarpX reference simulation, obtained analytically); the lens strengths used in the \texttt{ImpactX}+surrogate simulation, obtained analytically; and the optimal lens strengths that minimize emittance in the \texttt{ImpactX}+surrogate simulations.}
    \label{fig:lens_strengths}
\end{figure}

In this appendix, we discuss the optimal transport parameters found in the optimization workflow discussed in section \ref{sec:optimization}.
In Fig.~\ref{fig:lens_strengths}, we extend the plot in Fig.~\ref{fig:compare_warpx_impactx_surrogate} that shows lens strengths as calculated in the \verb+WarpX+ reference simulation and the \verb+ImpactX++surrogate simulation.
Fig.~\ref{fig:lens_strengths} includes the optimal lens strengths discovered by the workflow described in section \ref{sec:optimization} and that resulted in the beam parameters shown in figures~\ref{fig:optimized_emittance} and \ref{fig:optimized_width}.

Note that we present the lens strength in terms of magnetic field gradient.
The magnetic field of the lens circles around the $z$ axis and increases linearly with increasing radius.
Hence the field strength is expressed in terms of the slope of the field, $\frac{d\vec B}{dr}$, with units of Tesla per meter.
The magnetic field gradient relates to the lens strength $k$ shown in Fig.~\ref{fig:compare_warpx_impactx_surrogate} by the relation
\begin{equation}
    \frac{d\vec B}{dr}=\frac{\langle \gamma\rangle m_ec}{e}k^2,
\end{equation}
where $\langle \gamma \rangle$ is the average electron beam energy, $m_e$ is the electron mass, $c$ is the speed of light, and $e$ is the fundamental charge.
Presenting the lens strength in terms of the magnetic field gradient highlights the physical trend discovered by the optimization.

Observe how the lens parameters used in the \texttt{WarpX} and \texttt{ImpactX} +surrogate simulations oscillate around the optimal lens strengths. 
This oscillation in lens strengths corresponds with the oscillation of the beam widths and divergences observed in the \texttt{WarpX} simulation described in section \ref{sec:benchmark} and Fig.~\ref{fig:energy_emittance}.
That the oscillation is around the optimal lens parameters is another demonstration of the ability of the optimization to find a physical solution that improves the matching of the electron beam.
The smooth trend followed by the optimal lens strengths suggests a physical relation we have not yet derived, further highlighting the utility of this data-driven workflow.



\bibliographystyle{unsrt}
\bibliography{main}

\end{document}